\documentclass[aps,epsf,noshowpacs,superscriptaddress,pre]{revtex4}
\usepackage{bm}
\usepackage{graphicx}
\usepackage{color}
\usepackage{amscd}
\usepackage{epsfig}
\usepackage{amsmath}
\usepackage{amssymb}
\usepackage{enumerate}

\begin{document}

\title{Nonequilibrium Statistical Mechanics of Self-propelled Hard Rods}
\author{Aparna Baskaran}
\affiliation{Physics Department, Syracuse University, Syracuse, NY 13244, USA}

\author{M. Cristina Marchetti}
\affiliation{Physics Department, Syracuse University, Syracuse, NY 13244, USA}
\affiliation{Syracuse Biomaterials Institute, Syracuse University, Syracuse, NY 13244, USA}

\date{\today}

\begin{abstract}
Using tools of nonequilibirum mechanics, we study a model of self-propelled hard rods on a substrate in two dimensions to quantify the interplay of self-propulsion and excluded-volume effects. We derive of a Smoluchowski equation for the configurational probability density of self-propelled rods  that contains several modifications as compared to the familiar Smoluchowski equation for thermal rods.  As a side-product of our work, we also present a purely dynamical derivation of the Onsager form of
the mean field excluded volume interaction among thermal hard rods.
\end{abstract}

\maketitle

\address{Department of Physics, Syracuse University, Syracuse NY 13244 USA.}

\section{Introduction}

Self-propelled particles draw energy from internal or external sources and
dissipate this energy by moving through the medium they inhabit. A wide
class of systems, including fish schools, bacterial colonies, and monolayers of
vibrated granular rods can be  described within this paradigm.
These systems exhibit rich collective behavior, such as nonequilibrium phase
transitions between disordered and ordered (possibly moving) states and
novel long-range correlations and have been the subject of extensive
theoretical \cite{TonerPRL,TonerRev,KruseEPJE2005}, numerical \cite{Orsogna,Chate2,Lipowsky} and experimental investigations  \cite{actinexpts,bactexpts,rodsexpt1,rodsexpt2} in recent years.

Self-propelled particles are elongated in shape and have a
self-replenishing momentum along one direction of their long body axis. They generally experience attractive and repulsive interactions, both of a direct nature and mediated by the medium. One generic interaction that is relevant to all self-propelled systems is the short-range repulsive interaction arising from the finite size of the self-propelled units. Our goal here is to understand the interplay of self-propulsion and short range repulsive interactions in controlling the collective dynamics of the system. To this end, we consider the simplest implementation
of short range steric repulsion, which is the hard
particle limit, and consider
a minimal model of self-propelled hard rods moving on a substrate in two dimensions. We will show that self-propulsion modifies the momentum exchanged by hard rods upon collision and the resulting mean-field excluded
volume interaction as compared to the Onsager result for
passive rods~\cite{Onsager,EdDoiBook}.

The simplest model  of equilibrium nematic liquid crystals is a collection of long, thin hard rods~\cite{EdDoiBook}. In the
overdamped regime and at low density, the dynamical properties of the system are described by a
Smoluchowski equation for the configurational probability density, $c({\bf r}_1,\hat{\bf u}_1,t)$, of finding a rod with center of mass at ${\bf r}_1$ and long axis oriented along the unit vector $\hat{\bf u}_1$ at time $t$, given by
\begin{equation}
\frac{\partial c}{\partial t}+\bm\nabla_{1}\cdot{\bf J}+{\cal R}_1 \cdot{\bf J}^R=0\;,\label{Smol0}
\end{equation}
where ${\cal R}_1=\hat{\bf u}_1\times\partial_{\hat{\bf u}_1}$ is a rotation operator and ${\bf J}$ and ${\bf J}^R$ are translational and rotational currents, respectively, given by
\begin{subequations}
\begin{gather}
J_\alpha=-D_{\alpha \beta }\left[\partial_{1\beta}c+\frac{1}{k_{B}T}\left( \partial_{1\beta }V_{ex}\right) c\right]\;,\label{JT}\\
{\bf J}^R=-D_{R} \left[{\cal R}_1 c+\frac{1}{k_{B}T}\left({\cal R}_1 V_{ex}\right)c\right]\;,\label{JR}
\end{gather}
\end{subequations}
with
\begin{equation}
D_{\alpha\beta}=D_\bot\delta_{\alpha\beta}+(D_\Vert-D_\bot)\hat{u}_{1\alpha}\hat{u}_{1\beta}
\end{equation}
 a diffusion tensor that incorporates the anisotropy of translational diffusion of elongated objects, with $D_\Vert>D_\bot$, and $D_R$ the rate of rotational diffusion. The currents given in Eqs.~\eqref{JT} and \eqref{JR} incorporate both diffusion and binary interactions. The latter are described by  a mean field excluded volume potential, $V_{ex}$, given by
\begin{equation}
V_{ex}=k_{B}T\int_{\mathbf{\hat{u}}_{2}}\int_{\bf s}\left| \mathbf{\hat{u}}_{1}\times
\mathbf{\hat{u}}_{2}\right|c\left( \mathbf{r}
_{1}+{\bf s },\mathbf{\hat{u}}_{2},t\right)\;,\label{Vex}
\end{equation}
with $m$ the mass of a rod, $T$ the temperature, and ${\bf s} ={\bf r}_2-{\bf r}_1$ the separation between the cnters of mass of the two rods when they are at contact (Fig.~\label{fig:ExVol}). If the thickness of the rods is negligible compared to their length, we can approximate  ${\bf s} \simeq
s_{1}\mathbf{\widehat{u}}_{1}-s_{2}\mathbf{\widehat{u}}_{2}$, where
$-\ell/2\leq s_i\leq \ell/2$ for $i=1,2$, parametrizes the position along the
$i$-th rod of length $\ell$ measured from its center of mass. The integral over the vector ${\bf s }$
spans the area excluded to rod $1$ by a second rod
with center of mass at $\mathbf{r}_2$, oriented in the direction $\mathbf{\widehat{u}}_{2}$.  The excluded volume potential represents
the second virial coefficient of the static structure factor and was
first derived by Onsager \cite{Onsager}.
\begin{figure}[tbp]
\begin{center}
{\includegraphics[height=11.0cm, width=11.0cm]{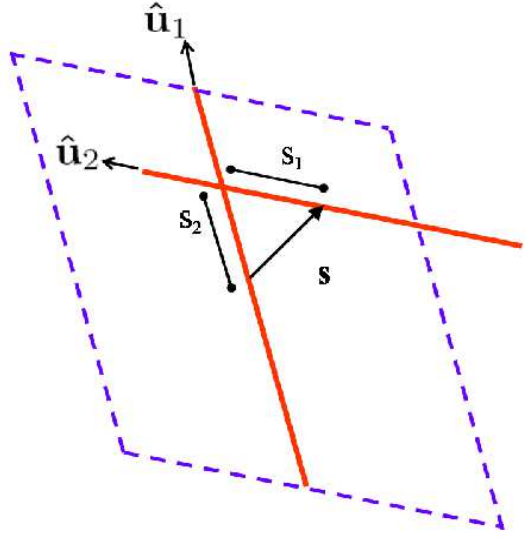}}
\end{center}
\caption{(color online) Long, thin rods of length $\ell$ are parametrized by the position ${\bf r}_i$ of their center of mass and a unit vector $\mathbf{\hat{u}}_{i}$ denoting the rod's orientation in the plane. The position along the $i$-th rod measured from the center of mass is denoted by $s_i$, with $-\ell/2\leq s_i\leq\ell/2$. The overlap situation shown in the figure requires
${\bf r}_1+ \mathbf{\hat{u}}_{1}s_1\simeq {\bf r}_2+ \mathbf{\hat{u}}_{2}s_2$, so that ${\bf s} =
s_{1}\mathbf{\widehat{u}}_{1}-s_{2}\mathbf{\widehat{u}}_{2}$ The area excluded by rod $2$ to rod  $1$ is the area of the dashed parallelogram spanned by the unit vectors $ \mathbf{\hat{u}}_{1}$ and $\mathbf{\hat{u}}_{2}$, given by $\ell^2|\mathbf{\hat{u}}_{1}\times \mathbf{\hat{u}}_{2}|$.}
\label{fig:ExVol}
\end{figure}

In this paper we present the derivation of a \emph{modified} Smoluchowski equation that describes the low density, overdamped dynamics of a collection of \emph{self-propelled} hard rods. We consider
long, thin hard rods moving on a substrate characterized by a friction constant
$\zeta $. Self propulsion is modeled by assuming  that each rod moves along one direction of its
long axis with a constant speed $v_{0}$. In addition, the rods experience binary hard-core collisions. This is the simplest model for a ``living nematic liquid crystal", a terminology that has been used to describe the collective behavior of a variety of intrinsically self-propelled systems, from bacterial suspensions to monolayers of vibrated granular rods. Since the rods
have a purely dynamical self-replenishing momentum, the
statistical mechanics needs to be derived from the underlying
trajectory dynamics. The details of the derivation are described in this paper. The outcome is a modified Smoluchowski  equation for the configurational probability density of the form given in Eq.~\eqref{Smol0}, but where the translational and rotational currents acquire  additional contributions due to self propulsion
and take the form
\begin{subequations}
\begin{gather}
J_\alpha=v_0\hat{u}_{1\alpha} c-D_{\alpha \beta }^{SP}\partial_{1\beta}c-\frac{D_{\alpha\beta}}{k_{B}T}\left( \partial_{1\beta }V_{ex}\right) c-\frac{D_\Vert m v_0^2}{2k_BT_a}I^{SP}_\alpha\;,\label{JTsp}\\
{\bf J}^R=-D_{R} \left[{\cal R}_1 c+\frac{1}{k_{B}T}\left({\cal R}_1 V_{ex}\right)c\right]-\frac{D_R m v_0^2}{2k_BT_a}{\bf I}_R^{SP}\;,\label{JRsp}
\end{gather}\label{JSP}
\end{subequations}
where
\begin{eqnarray}
D_{\alpha \beta }^{SP}&&=D_{\alpha \beta }+D_S~\hat{u}_{1\alpha}\hat{u}_{1\beta}\nonumber\\
&&=D_\bot\delta_{\alpha\beta}+(D_\Vert+D_S-D_\bot)\hat{u}_{1\alpha}\hat{u}_{1\beta}\;,\label{DSP}
\end{eqnarray}
with $D_S=v^{2}_{0}/\zeta$, is the diffusion tensor.
Self-propulsion modifies  the familiar Smoluchowski equation for hard rods in several important ways.
The first
modification is the convective mass flux at the self-propulsion
speed $v_{0}$ along the axis of the rod, described by the first term on the right hand side of Eq.~\eqref{JTsp}.
Secondly, self-propulsion enhances the longitudinal diffusion
constant $D_{\parallel }$ of the rods, according to $D_{\parallel
}\rightarrow D_{\parallel }+D_S=D_{\parallel }(1+mv_{0}^{2}/k_{B}T_a)$, as shown in eq.~\eqref{DSP} This
enhancement arises because self-propelled particles perform a
\emph{persistent random walk}, as recently pointed out by other
authors \cite{PersistentRW1,PersistentRW2}. Finally,  the momentum exchanged by
two rods upon collision is rendered highly anisotropic by
self-propulsion. This yields the additional collisional contributions to the excluded
volume interaction described by the last terms in Eqs.~\eqref{JTsp} and \eqref{JRsp}. The precise form of these contributions can be found in Section IV.B.
The Smoluchowski equation for self-propelled hard rods is the central result of this work. It has also been shown by us that the novel terms arising from self-propulsion have important consequences for
the long-wavelength, long-time behavior of the system by introducing new terms in the coarse-grained equations for
the dynamics of conserved quantities and  broken symmetry
variables. These hydrodynamic signatures have been reported in
earlier work \cite{ABMCM_PRL}.
Finally, an additional result of the work presented here
is a
purely dynamical derivation of the familiar Onsager excluded
volume potential for equilibrium hard rods, given in Section IV.A.

The layout of this paper is as follows. In Section II we analyze the
trajectory of hard rods moving on a substrate in two dimensions. Using the fact that
their trajectories are piece-wise differentiable, with singularities at the
time of each  collision, we derive an expression for the collision operator governing the momentum exchanged in a binary collision. In Section III we derive a formal hierarchy of Fokker-Planck
equations governing the noise-averaged dynamics of a collections of self-propelled hard rods.
In section IV  we consider the limit of high friction with the substrate that yields a fast relaxation of the linear and angular momentum degrees of freedom, relative to that of the configurational degrees of freedom. This approximation, together with a low density closure of the Fokker-Planck hierarchy, allows us to  derive the Smoluchowski equation. We conclude with a brief discussion.

\section{Binary Collision of Hard Rods}

\begin{figure}[tbp]
\begin{center}
{\includegraphics[height=5.0cm, width=9.0cm]{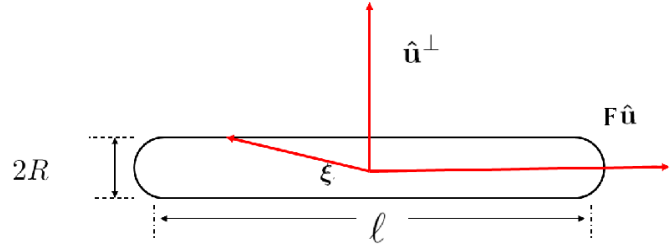}}
\end{center}
\caption{(color online) Each self-propelled particle is  modeled as a capped rectangle of uniform mass density, consisting of a rectangle of length $\ell $ and thickness $
2R$ capped at the two short sides by semicircles of radius $R$ (the width of the rod is exaggerated for clarity). Self-propulsion is provided by a force of constant magnitude $F$ directed along the "head" of the rod. The vector  $\bm \xi$ locates points on the boundary of the rod relative to the center of mass. The unit vector $\mathbf{\widehat{u}}_\bot$ is defined as $\mathbf{\widehat{u}}_\bot=\mathbf{\widehat{z}}\times\mathbf{\widehat{u}}$, where $\mathbf{\widehat{z}}$ points out of the plane.}
\label{fig:onerod}
\end{figure}

Our model is a collection of
self-propelled  thin hard rods of length $\ell $
and mass $m$, confined to two dimensions. Although we will focus below on the limit of long, thin rods, to describe a binary collision  we need to incorporate their finite thickness of the rods. We model each rod as  a capped rectangle of uniform mass density, consisting of a rectangle of length $\ell $ and thickness $%
2R\ll\ell$, capped at the two short sides by semicircles of radius $R$, as shown in Fig. (\ref{fig:onerod}). Each rod  is
described by the position $\mathbf{r}$ of its center of mass and a unit vector $\mathbf{\hat{u}}$ $=\cos \theta \hat{\mathbf x}+\sin
\theta \hat{\mathbf y}$ directed along its long axis. The rods have head/tail, i.e., nematic symmetry. This symmetry is
broken by self-propulsion that is
implemented by assuming that a force ${\bf F}$ of constant magnitude  and directed along the rod's long axis acts on the center of mass of
each rod. The direction of the self-propulsion force will be referred to as  the ''head'' of the rod and the unit vector $\mathbf{\hat u}$ is chosen to point in the direction of the head, so that
${\bf F}=F\hat{\bf u}$.

The rods move on a passive medium that provides frictional
damping to their motion. Any other physical or chemical process
that may be present in the system is assumed to occur on a fast
time scale, such that it can be modeled as an additive
Markovian white noise. The dynamics of a  self propelled
rod is then described
by Langevin equations for the center of mass velocity $\mathbf{v}=\partial _{t}
\mathbf{r}$ and the angular velocity $\bm\omega=\mathbf{\hat{z}}\omega=\mathbf{\hat{z}}\partial _{t}\theta $,
where $\mathbf{\hat z}$ is normal to the plane of motion.
The equations of motion are given by
\begin{subequations}
\begin{gather}
m \partial _{t}v_{\alpha }=-\zeta _{\alpha \beta }v_{\beta }+F\hat{u}_{\alpha
}+\eta _{\alpha }\left( t\right)\;,\\
\partial _{t}{\omega }=-\zeta ^{R}{%
\omega }+{\eta }_{R}\left( t\right) ,  \label{2.1}
\end{gather}
\end{subequations}
where  $\zeta _{\alpha \beta }=\zeta _{\Vert }\hat{u}%
_{\alpha }\hat{u}_{\beta }+\zeta _{\bot }\left( \delta _{\alpha \beta }-\hat{%
u}_{\alpha }\hat{u}_{\beta }\right) $ is a translational friction tensor
with $\zeta _{\Vert }<\zeta _{\bot }$ reflecting the fact that frictional
damping is smaller for motion along the long axis of the rod,
$\zeta _{R}$ is a rotational friction constant, and $%
\bm\eta $ and $\eta_{R}$ are white noise terms with zero mean and correlations
\begin{subequations}
\begin{gather}
\left\langle \eta _{\alpha }\left( t\right) \eta _{\beta }\left(
t^{\prime }\right) \right\rangle =\Delta _{\alpha \beta }(\mathbf{\hat u})~\delta \left(
t-t^{\prime }\right)\;,\label{2.2a}\\
\left\langle \eta _{R}\left( t\right) \eta
_{R}\left( t^{\prime }\right) \right\rangle =\Delta _{R}~\delta \left( t-t^{\prime }\right)\;.  \label{2.2}
\end{gather}
\end{subequations}
For
simplicity, we assume that the noise amplitudes $\Delta_{\alpha\beta}$ and $\Delta _{R}$ have the same form as in equilibrium,
\begin{subequations}
\begin{gather}
\Delta _{\alpha \beta }(\mathbf{\hat u})=2k_{B}T_{a}\zeta _{\alpha \beta }(\mathbf{\hat u})/m\;,\\
\Delta _{R}=2k_{B}T_{a}\zeta_{R}/I\;,
\end{gather}
\end{subequations}
with $T_{a}$  an active temperature and $I$ the component of the moment of
inertia tensor ${\bf I}$ of the rod along the long axis,
\begin{equation}
\label{I}
I=\frac{m}{2\ell +\pi R}\left[ \frac{2\ell }{3}
\left( 3R^{2}+\frac{\ell ^{2}}{4}\right) +\pi R\left( \frac{R^{2}}{2}
+\frac{\ell ^{2}}{4}\right) \right] \sim \frac{m\ell ^{2}}{12}\;.
\end{equation}
The last approximate equality in Eq.~\eqref{I} holds in the limits $\ell\gg 2R$ of long, thin rods.  The active temperature $T_{a}$ is generally different from the
thermodynamic temperature of the system and is  a measure of the noise
amplitude.

\begin{figure}[tbp]
\begin{center}
{\includegraphics[height=9.0cm, width=9.0cm]{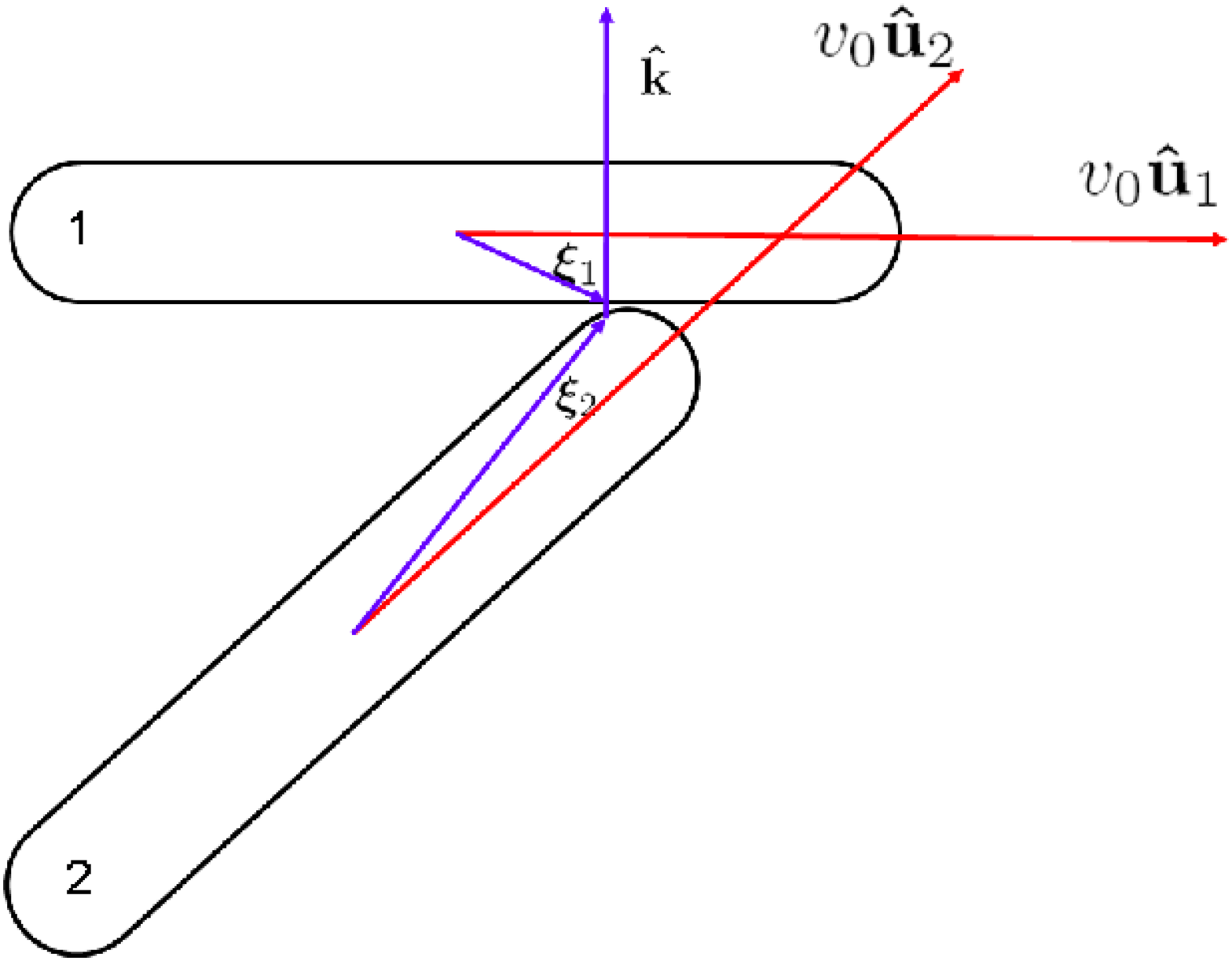}}
\end{center}
\caption{(color online) A cap-to-side collision of two self-propelled hard rods
(the width of the rod is exaggerated for clarity). $\widehat{\bf k}$ is a
unit vector from rod 2 to rod 1  normal to the point
of contact. Points on the side of the rods are identified by
vectors ${\bm {\xi}}_i$. } \label{fig:collision}

\end{figure}

The rods interact with each other exclusively via hard-core interactions. The collisions are instantaneous and conserve energy and momentum of the colliding rods. To incorporate these
interactions in the Langevin equations for the particles, we
need to construct a collision operator that generates the
instantaneous collision. We consider two rods and denote by $t=0$ the origin of time. The two rods travel freely with linear and angular velocities $x_1=({\bf v}_1,\bm\omega_1)$ and $x_2=({\bf v}_2,\bm\omega_2)$  until a time $\tau(\Gamma_1,\Gamma_2)$ when they come into
contact, where $\Gamma_i =\{\mathbf{r}_{i},\mathbf{\hat{u}}_{i},\mathbf{v}_{i},{\bm
\omega}\}$ is the phase point of each rod.
Denoting by a prime the post-collisional velocities,
$x^\prime_i=({\bf v}_i^\prime,\bm\omega_i^\prime)_{i=1,2}$, the time dependence
of the observables $x_{i}$ can be written as
\begin{equation}
x_{i}\left( t\right) =x_{i}\Theta \left( \tau \left( \Gamma_1,\Gamma_2 \right)
-t\right) +x_{i}^{\prime }\Theta \left( t-\tau \left(  \Gamma_1,\Gamma_2 \right) \right)\;.
\label{2.4a}
\end{equation}
The
equation of motion for  $x_{i}$ is then
\begin{equation}
\partial _{t}x_{i}=\Delta x_{i}~\delta \left( t-\tau \left( \Gamma_1,\Gamma_2 \right)
\right) \;,  \label{xdot}
\end{equation}
with $\Delta x_{i}=x_{i}-x_{i}^{\prime }$.
The post-collisional velocities are easily calculated by imposing conservation of energy and of linear and angular momentum, ${\bf p}_i=m{\bf v}_i$ and ${\bf L}_i=I\bm\omega_i$, with the result~\cite{Huthmann1999,Otto2006}
\begin{subequations}
\begin{gather}
\mathbf{p}_{i}^{\prime }=\mathbf{p}_{i}-\mathbf{\hat{k}}_{ji}A\;,\\
\mathbf{L}_{i}^{\prime }=\mathbf{L}_{i}-{\bm\xi }_{i}\times \mathbf{\hat{k}}_{ji}A\;,
\label{2.3}
\end{gather}
\end{subequations}
where  $\widehat{\mathbf{
k}}_{ji}$ is the unit normal at the point of contact of the two rods
and is directed from rod $j$ to rod $i$, and $\bm\xi _{i}$
is the vector  from the center of mass of rod $i$ to the point of
contact of the two rods, as shown in Fig.~\ref{fig:collision}.
The magnitude of the momentum transfer $A$ is
\begin{equation}
A=\frac{m\mathbf{\hat{k}}_{21}\cdot \mathbf{V}_{12}}{1+\frac{m}{2}\left( {\bm%
\xi }_{1}\times \mathbf{\hat{k}}_{21}\right) \cdot {\bf I}^{-1}\cdot \left( {\bm%
\xi }_{1}\times \mathbf{\hat{k}}_{21}\right) \mathbf{+}\frac{m}{2}\left( {\bm%
\xi }_{2}\times \mathbf{\hat{k}}_{21}\right) \cdot {\bf I}^{-1}\cdot \left( {\bm%
\xi }_{2}\times \mathbf{\hat{k}}_{21}\right) }  \label{A}
\end{equation}
with $\mathbf{V}_{12}=\mathbf{v}_{12}+{\bm\omega }_{1}\times {\bm\xi }%
_{1}-{\bm\omega }_{2}\times {\bm\xi }_{2} $ the relative
velocity of the two rods at the point of contact, and ${\bf v}_{12}={\bf v}_1-{\bf v}_2$.  To render Eq.~\eqref{xdot} explicit we need an expression for the time $\tau(\Gamma_1,\Gamma_2)$.
The condition of contact can be written as the requirement that
$\mathbf{r}_{1}(\tau)+$ ${\bm\xi }_{1}(\tau)=\mathbf{r}_{2}(\tau)+{\bm\xi }_{2}(\tau)$
for some value of $\bm\xi_1$ and $\bm\xi_2$ on the surface of the two rods,
where $\mathbf{r}_{i}\left(
\tau \right) =\mathbf{r}_{i}+\mathbf{v}_{i}\tau $ and
 $\mathbf{\hat{u}}
_{i}\left( \tau \right) =\cos \left( \theta _{i}+\omega _{i}\tau \right)
\hat{\bf x}+\sin \left( \theta _{i}+\omega _{i}\tau \right) \hat{\bf y}$.
It is apparent from Fig.~\ref{fig:collision}   that $%
\mathbf{r}_{12}+{\bm\xi }_{1}{-\bm\xi }_{2}$ must lie along $\mathbf{\hat{k}}_{21}
$. The condition of contact can then be written as two scalar equations, given by
\begin{subequations}
\begin{gather}
\left( \mathbf{r}_{12}+{\bm\xi }_{1}{-\bm\xi }_{2}\right) \cdot \mathbf{\hat{%
k}}_{21}=\left| \mathbf{r}_{12}+{\bm\xi }_{1}{-\bm\xi }_{2}\right| =0\;,\label{contact1}\\
\left(
\mathbf{r}_{12}+{\bm\xi }_{1}{-\bm\xi }_{2}\right) \cdot \left( \mathbf{\hat{%
z}}\times \mathbf{\hat{k}}_{21}\right) =0\;,  \label{contact2}
\end{gather}
\end{subequations}
where all variables are evaluated at time $\tau$. Equations~\eqref{contact1} and \eqref{contact2} are implicit equations for $\tau(\Gamma_1,\Gamma_2)$.
The first  condition, Eq.~\eqref{contact1}, imposes that two rods  be in contact at any point along their surface. The second condition, Eq.~\eqref{contact2},  determines the precise point on the surface of each rod.  As an illustration we consider  the collision
 shown in Fig.~\ref{fig:collision} when the cap
of rod $2$ comes into contact with a side of rod $1$. In this case, $%
\mathbf{\hat{k}}_{12}\equiv\mathbf{\hat{u}}_{1}^{\bot }=\mathbf{\hat{z}}\times
\mathbf{\hat{u}}_{1}$ and the surface of the rod 1 is parametrized by $\bm{\xi }_{1}=s_{1}\mathbf{\hat{u}}_{1}-R\mathbf{%
\hat{u}}_{1}^{\bot }$, with $\mathbf{\hat{u}}_i^\bot=\mathbf{\hat{z}}\times\mathbf{\hat{u}}_i$. The point of contact on rod 2 is simply $\bm{\xi }_{2}=\frac{\ell }{2}\mathbf{\hat{u}}_{2}+R\mathbf{\hat{u}}_{1}^{\bot }$.
The contact conditions Eqs.~\eqref{contact1} and \eqref{contact2} become
\begin{subequations}
\begin{gather}
\mathbf{r}_{12}\cdot \mathbf{\hat{u}}_{1}^{\bot }-\frac{\ell }{2}\mathbf{%
\hat{u}}_{2}\cdot \mathbf{\hat{u}}_{1}^{\bot }-2R=0\;,  \label{contact1s}\\
\mathbf{r}_{12}\cdot \mathbf{\hat{u}}_{1}-\frac{\ell }{2}\mathbf{\hat{u}}%
_{2}\cdot \mathbf{\hat{u}}_{1}+s_{1}=0\;.  \label{contact2s}
\end{gather}
\end{subequations}
Eq.~\eqref{contact1s} requires the rods to be at contact at any point along the side of rod $1$, while Eq.~\eqref{contact2s} determines the position of contact along the side of  rod
$1$.

To obtain an expression that can be used to eliminate $\tau(\Gamma_1,\Gamma_2)$ from the equation of motion, Eq.~(\ref{xdot}) we use the identity
\begin{equation}
\delta \left( f\left( x\right) -f_{0}\right) =\frac{\delta \left(
x-x_{0}\right) }{\left| \frac{\partial f}{\partial x}|_{x_{0}}\right| }\;.
\label{identity}
\end{equation}
Using Eq.~\eqref{identity}, together with
\begin{equation}
\frac{\partial }{\partial t}\left( \mathbf{r}_{12}+{\bm\xi }_{1}{-\bm\xi }%
_{2}\right) \cdot \mathbf{\hat{k}}_{21}|_{t=\tau }=\mathbf{V}_{12}\left(
\tau \right) \cdot \mathbf{\hat{k}}_{21}  \label{2.5.6}
\end{equation}
we can rewrite the temporal $\delta $ function in the
equation of motion, Eq.~\eqref{xdot}  as
\begin{equation}
\delta \left( t-\tau \left( \Gamma_1,\Gamma_2 \right) \right) =\delta \left( \left(
\mathbf{r}_{12}+{\bm\xi }_{1}{-\bm\xi }_{2}\right) \cdot \mathbf{\hat{k}}%
_{21}\right) \left| \mathbf{V}_{12}\cdot \mathbf{\hat{k}}_{21}\right|\;.
\label{2.5.7}
\end{equation}
Finally, although Eqs. (\ref{contact1}-\ref{contact2}) determine the
contact condition, an actual collision results from the contact only if
the pre-collisional velocities at the point of contact are such that
the two particles are
moving towards each other, i.e., $\mathbf{V}_{12}\cdot \mathbf{\hat{k}}%
_{21}<0$. Putting all of these results  together,  the equation of motion
for the observable $x_{i}$ can be written in the form
\begin{equation}
\partial _{t}x_{i}=T\left( 1,2\right) x_{i}  \label{2.6.1}\;,
\end{equation}
where $T(1,2)$ is a binary collision operator, given by
\begin{equation}
T\left( 1,2\right) =\int_{\bm\xi _{1},\bm\xi _{2}}\int_{\mathbf{\hat{k}}_{21}}\Theta \left( -%
\mathbf{V}_{12}\cdot \mathbf{\hat{k}}_{21}\right) \left| \mathbf{V}%
_{12}\cdot \mathbf{\hat{k}}_{21}\right| \delta \left( \left( \mathbf{r}_{12}+%
{\bm\xi }_{1}{-\bm\xi }_{2}\right) \cdot \mathbf{\hat{k}}_{21}\right) \delta
\left( \left( \mathbf{r}_{12}+{\bm\xi }_{1}{-\bm\xi }_{2}\right) \cdot
\left( \mathbf{\hat{z}}\times \mathbf{\hat{k}}_{21}\right) \right) \left(
\hat{b}_{12}-1\right)\;,   \label{T12}
\end{equation}
with $\hat{b}_{12}$ a substitution operator such that $\hat{b}_{12}x_{i}=x_{i}^{\prime }$.
The integration in Eq.~\eqref{T12} ranges over all physical collision geometries. In the following,
we will focus on the case of rods of large aspect ratio ($\ell\gg 2R$). In this limit, cap-on-cap collisions are rare relative to to cap-on-side collisions and will be
neglected. For cap-on-side collisions, the $\int_{\bm\xi _{1},\bm\xi _{2}}\int_{\mathbf{\hat{k}}_{21}}$  can be given an explicit representation of the
form
\begin{eqnarray}
\int_{\bm\xi_{1},\bm\xi_{2}}\int_{\mathbf{\hat{k}}_{21}}...=&&\int_{s_{1},s_{2}}\int_{\mathbf{\hat{k}}_{21}}...~
\Big\{\Big[ \delta \left( s_{2}+\frac{\ell }{2}\right) +\delta \left( s_{2}-\frac{\ell }{2}\right) \Big]
\Big[ \delta \left( \mathbf{\hat{k}}_{21}-\mathbf{\hat{u}}_{1}^{\bot }\right) +\delta \left( \mathbf{\hat{k}}_{21}+\mathbf{\hat{u}}_{1}^{\bot }\right) \Big] \nonumber \\
&&
+\left[ \delta \left( s_{1}+\frac{\ell }{2}\right) +\delta \left( s_{1}-%
\frac{\ell }{2}\right) \right] \left[ \delta \left( \mathbf{\hat{k}}_{21}-%
\mathbf{\hat{u}}_{2}^{\bot }\right) +\delta \left( \mathbf{\hat{k}}_{21}+%
\mathbf{\hat{u}}_{2}^{\bot }\right) \right]\Big\}\nonumber\\
&&\equiv \int_{s_{1},s_{2}}\int_{\mathbf{\hat{k}}_{21}}...~
\delta
\left( \Gamma _{cont}( s_{1},s_{2},\widehat{\mathbf{k}}_{21}) \right)
\;. \label{int}
\end{eqnarray}
where $\int_s...=\int_{-\ell/2}^{\ell/2}ds...$  and  the last line simply defines a compact notation for the condition of contact. Equation~\eqref{T12}, with the expression given in Eq.~\eqref{int} for the range of integration, is the generator of collisional dynamics that will be used in the rest of this work.

The above considerations are
readily generalized to a system of $N$ rods by considering only binary collisions since collisions among particles with hard core interactions are instantaneous and the probability of three or
more particles being at contact at the same instant is of measure zero. In addition, the dynamics arising from the white noise due to the interaction with the substrate  is described by a continuous generator and does not lead to any additional singularity.
The derivation of the collision trajectory can then be carried out as before. The dynamics of a system of $N$
self-propelled hard rods moving on a passive substrate  in the $N$-rod phase
space $\Gamma=\left\{\Gamma_1,\Gamma_2,...,\Gamma_N\right\}$ is controlled by a set of coupled Langevin equations for the linear velocities $\partial _{t}\mathbf{r}%
_{i}=\mathbf{v}_{i}$ and angular velocities ${\bm\omega }%
_{i}=\mathbf{\hat{u}}_{i}\times \partial _{t}\mathbf{\hat{u}}_{i}= \mathbf{\hat z}\omega_i$, given by
\begin{subequations}
\begin{gather}
m \partial _{t}\mathbf{v}_{i}=m\sum_{j\neq i}T\left( i,j\right) \mathbf{v}_{i}-%
{{\bm\zeta }}_{i}\cdot \mathbf{v}_{i}+F\mathbf{\hat{u}}%
_{i}+{\bm\eta }_{i}  \label{Langevin-v}\\
\partial _{t}{\omega }_{i}=\sum_{j\neq i}T\left( i,j\right) {\omega }%
_{i}-\zeta _{R}{\omega }_{i}+{\eta }_{i}^{R}  \label{Langevin-om}
\end{gather}
\end{subequations}
where ${{\bm\zeta }}_{i}$ is a friction tensor with components
$\zeta _{i,\alpha \beta } =(\zeta _{\Vert }-\zeta_{\bot })\hat{u}_{i\alpha } \hat{u}_{i\beta }+\zeta_{\bot } \delta _{\alpha \beta } $, ${\bm\eta }_{i}$ and $\eta _{i}^{R}$ are Gaussian random forces with zero mean and correlations as defined in Eqs.~(\ref{2.2a}-\ref{2.2}). Noise sources associated with different values of the rod index $i$ are uncorrelated.

\section{Non-equilibrium Statistical Mechanics}

We are interested in the collective behavior  of self-propelled rods in the limit of large friction with the substrate. In this limit one can consider a description that applies on time scales
$t\gg m \zeta_\Vert^{-1}$ that neglects the fast
dynamics of the linear and angular velocities and only considers the dynamics of the coordinate degrees of freedom,  $\left\{
\mathbf{r}_{i},\mathbf{\hat{u}}_{i}\right\} _{i=1}^{N}$.
The derivation of the overdamped dynamics must, however, be carried out carefully  when dealing with the singular limit of hard particles, where the
interactions, even though conservative, depend on the velocities of the
particles.

With this goal in mind, we consider an observable of the system $\hat{A}\left( \Gamma \right) $, where $
\Gamma $ is an N-rod phase point.
Using Ito calculus \cite
{OsendalBook},  the stochastic equation of motion for the
observable $\hat{A}$ can be derived from the equations
of motion, Eq.~\eqref{Langevin-v} and \eqref{Langevin-om},  for the phase space variables,
with the result
\begin{equation}
\frac{\partial \hat{A}\left( \Gamma ,t\right) }{\partial t}=\hat{L}\hat{A}%
\left( \Gamma ,t\right)\;,   \label{Aeq}
\end{equation}
where the operator $\hat{L}$ is given by
\begin{eqnarray}
\hat{L} &=&\sum_{i=1}^{N}\Big\{\mathbf{v}_{i}\cdot \nabla _{\mathbf{r}_{i}}+{\bm
\omega }_{i}\cdot{\cal R}_i +\frac{F}{m}\widehat{\mathbf{u}}_{i
}\cdot\bm\nabla _{{\bf v}_{i }}
-\frac{1}{m}\zeta _{\alpha \beta }^{i}v_{i\beta }\partial _{v_{i\alpha }}-\zeta
^{R}\omega _{i }\partial _{\omega _{i }}  \nonumber \\
&&+\frac{1}{m}\bm\eta _i\cdot\bm\nabla_{{\bf v}_{i }}+\eta _{i }^{R}\partial
_{\omega _{i }}-\frac{1}{2m}\Delta _{\alpha \beta }^{iT}\partial _{v_{i\alpha
}}\partial _{v_{i\beta }}
-\frac{1}{2}\Delta ^{R}\partial _{\omega _{i}}^{2}f \Big\}+\frac{1}{2}\sum_{i,j\neq
i}T\left( i,j\right)\;,  \label{Lhat}
\end{eqnarray}
where ${\cal R}_i= \mathbf{\hat{u}}_{i}\times \partial_{\mathbf{\hat{u}}_{i}}$. The binary substitution operator $\hat{b}_{ij}$ contained in $T(i,j)\equiv T(\Gamma_i,\Gamma_j)$ replaces the velocities of the $(i,j)$ pair with their post-collisional values, according to
$\hat{b}_{ij}\hat{A}\left( \mathbf{v}_{i},\mathbf{v}_{j},{\bm\omega }_{i},{\bm\omega }%
_{j}\right) =\hat{A}\left( \mathbf{v}_{i}^{\prime },\mathbf{v}_{j}^{\prime },{\bm%
\omega }_{i}^{\prime },{\bm\omega }_{j}^{\prime }\right) $, leaving the velocities of all other particles unchanged.
Equation \eqref{Aeq} describes the stochastic dynamics of any observable $\hat{A}(\Gamma)$ for
given initial conditions in phase space. The quantity of interest here, however,
is not the stochastic observable itself,  but rather its ensemble averaged
value for a given ensemble of initial conditions,  $\hat{\rho}_{N}\left(
\Gamma \right) $, i.e.,
\begin{equation}
\left\langle \hat{A}\left( t\right) \right\rangle _{ens}=\int d\Gamma \hat{%
\rho}_{N}\left( \Gamma \right) \hat{A}\left( \Gamma ,t\right)\;.   \label{Aens1}
\end{equation}
Equivalently, we can treat the phase space density as the dynamical quantity
and write  the ensemble average of an observable as
\begin{equation}
\left\langle \hat{A}\left( t\right) \right\rangle _{ens}=\int d\Gamma \hat{%
\rho}_{N}\left( \Gamma ,t\right) \hat{A}\left( \Gamma \right)\;.  \label{Aens2}
\end{equation}
Equation \eqref{Aens2}  \textit{defines} the dynamics of the phase
space probability density, $\hat{\rho}_{N}(\Gamma,t)$. Taking the time
derivative of both Eqs. \eqref{Aens1} and \eqref{Aens2} and defining an adjoint operator $\hat{\cal L}$,
\begin{eqnarray}
\int d\Gamma~ \partial _{t}\hat{\rho}\left( \Gamma ,t\right) A\left( \Gamma
\right)  &=&\int d\Gamma ~\hat{\rho}_{N}\left( \Gamma \right) \partial _{t}%
\hat{A}\left( \Gamma ,t\right)   \nonumber \\
&=&\int d\Gamma ~\hat{\rho}_{N}\left( \Gamma \right) \hat{L}\hat{A}\left(
\Gamma ,t\right)   \nonumber \\
&=&-\int d\Gamma~ \left[ \hat{\cal{L}}\hat{\rho}_{N}\left( \Gamma ,t\right)\right]
\hat{A}\left( \Gamma \right)\;,
\end{eqnarray}
we obtain a Liouville-like equation describing the time evolution of of the phase space probability density,
\begin{equation}
\left( \frac{\partial }{\partial t}+\mathcal{\hat L}\right) \rho _{N}\left(
\Gamma ,t\right) =0 \;, \label{Liouville}
\end{equation}
where
\begin{eqnarray}
\hat{L} &=&\sum_{i=1}^{N}\Big\{\mathbf{v}_{i}\cdot \nabla _{\mathbf{r}_{i}}+{\bm
\omega }_{i}\cdot{\cal R}_i +\frac{F}{m}\widehat{\mathbf{u}}_{i
}\cdot\bm\nabla _{{\bf v}_{i }}
-\frac{1}{m}\zeta _{\alpha \beta }^{i}v_{i\beta }\partial _{v_{i\alpha }}-\zeta
^{R}\omega _{i }\partial _{\omega _{i }}  \nonumber \\
&&+\frac{1}{m}\bm\eta _i\cdot\bm\nabla_{{\bf v}_{i }}+\eta _{i }^{R}\partial
_{\omega _{i }}-\frac{1}{2m}\Delta _{\alpha \beta }^{iT}\partial _{v_{i\alpha
}}\partial _{v_{i\beta }}
-\frac{1}{2}\Delta ^{R}\partial _{\omega _{i}}^{2}f \Big\}-\frac{1}{2}\sum_{i,j\neq
i}\bar{T}\left( i,j\right)\;.\label{Lcal}
\end{eqnarray}
The single-particle part of $\hat{\mathcal{L}}$ is identified by a simple integration by parts.
To determine the binary collision operator $\bar{T}(i,j)$
one needs to explicitly construct the restituting collision, with the result
\begin{eqnarray}
\overline{T}\left( 1,2\right) \hat{\rho} &=&\int_{s_{1}s_{2}}\int_{\hat{\bf k}}\delta
\left( \Gamma _{cont}\left( s_{1},s_{2},\widehat{\mathbf{k}}\right) \right)
\left( \hat{b}_{12}^{-1}-1\right)
 \left| \mathbf{V}_{12}\cdot \widehat{\mathbf{k}}\right|
\Theta \left( -\mathbf{V}_{12}\cdot \widehat{\mathbf{k}}\right) \hat{\rho}%
\;,  \label{Tbar}
\end{eqnarray}
where $\hat{b}_{ij}^{-1}$ is the generator of restituting collisions, i.e., $%
\hat{b}_{ij}^{-1}A\left( x_i^{\prime},x_j^{\prime }\right) =A\left(x_i,x_j\right) $,
i.e., it replaces the post-collisional velocities of the pair with their pre-collisional values, and
$\delta \left( \Gamma _{cont}\left(
s_{1},s_{2},\widehat{\mathbf{k}}\right) \right) $, defined in
Eq.~\eqref{int}, enforces the condition of contact.

Finally, we average over the noise and define $\rho
_{N}=\left\langle \hat{\rho}\right\rangle $. The dynamical
equation describing the evolution of the noise-averaged density is
\begin{equation}
\left( \frac{\partial }{\partial t}+\mathcal{L}\right) \rho _{N}\left(
\Gamma ,t\right) =0\;,  \label{LFP}
\end{equation}
where $\mathcal{L}$ is the Liouville Fokker Planck operator, given by
\begin{eqnarray}
\mathcal{L} &=&\sum_{i=1}^{N}\Big\{\mathbf{v}_{i}\cdot \nabla _{\mathbf{r}_{i}}+{%
\bm\omega }_{i}\cdot \left( \mathbf{\hat{u}}_{i}\times \frac{\partial }{%
\partial \mathbf{\hat{u}}_{i}}\right) +\frac{F}{m}\widehat{\mathbf{u}}_{i
}\cdot\bm\nabla_{{\bf v}_{i }}
-\frac{\zeta _{\alpha \beta }^{i}}{m}\partial _{v_{i\alpha }}v_{i\beta }-\zeta
^{R}\partial _{\omega _{i }}\omega _{i }  \nonumber \\
&&-\frac{1}{2m}\Delta _{\alpha \beta }^{iT}\partial _{v_{i\alpha }}\partial _{v_{i\beta
}}-\frac{1}{2}\Delta ^{R}\partial _{\omega _{i}}^{2}f\Big\}-\frac{1}{2}\sum_{i,j\neq i}%
\overline{T}\left( i,j\right)\;.   \label{B.1.10}
\end{eqnarray}

The formulation just described is exact and can be used, for instance, to evaluate time
correlation functions for the system.  To proceed, we will now restrict our attention to a low-density collections of self-propelled rods. In this case one can make systematic approximations to obtain the effective coarse-grained theory in the form of a Smoluchowski equation.

\section{Low Density Effective Statistical Mechanics}

In order to carry out systematic approximations in the low density limit, it is convenient to define a hierarchy of reduced distribution
functions as
\begin{equation}
f_m \left( \Gamma_1,...,\Gamma_{m},t\right) =V^{m}\int
d\Gamma _{m+1}... d\Gamma _{N}~\rho _{N}\left( \Gamma ,t\right) \;.
\end{equation}
The Liouville Fokker-Planck equation, Eq.~\eqref{LFP}, can then be rewritten as an infinite hierarchy of
equations for the reduced distribution functions. The $m$-th equation for  $f_m$ couples to $%
f_{m+1}$. The resulting Fokker-Planck hierarchy is analogous to the BBGKY hierarchy for Hamiltonian
systems and forms the starting point for
constructing approximate theories to describe the system in various regimes
of interest.

At low density, we consider the first equation of the hierarchy for the one-particle distribution  function $%
f_1\left( \Gamma_{1},t\right)$, given by
\begin{eqnarray}
\frac{\partial f_1(\Gamma_1,t)}{\partial t}+\mathcal{D}f_1(\Gamma_1,t) &=&\int d\Gamma_{2}~\overline{T}\left( 1,2\right) f_2
(\Gamma_1,\Gamma_2,t)\;,    \label{FP}
\end{eqnarray}
where the one-particle operator $\mathcal{D}$ is given by
\begin{equation}
\mathcal{D}f_1=\mathbf{v}_{1}\cdot \nabla _{\mathbf{r}%
_{1}}f_1+{\bm\omega }_{1}\cdot{\cal R}_1 f_1+\frac{F}{m}\widehat{\mathbf{u}%
_{1 }}\cdot\bm\nabla _{{\bf v}_{1}}f_1-\frac{1}{m}\zeta _{\alpha \beta }^{1}\partial
_{v_{1\alpha }}\left( v_{1\beta }f_1\right) -\zeta ^{R}\partial _{\omega %
_{1}} \left( {\omega }_{1}f_1\right) -\frac{1}{2m}\Delta _{\alpha \beta
}^{1T}\partial _{v_{1\alpha }}\partial _{v_{1\beta }}f_1-\frac{1}{2}\Delta ^{R}\partial
_{\omega _{1}}^{2}f_1\;.
\end{equation}
Equation \eqref{FP} is a generalized Fokker-Planck equation that includes binary collisions. We are interested in the limit of large friction, where linear and angular velocities relax on fast time scales. Our goal is to construct an approximate closed equation for a local concentration field $c\left( \mathbf{r},\mathbf{\hat{u}}
,t\right) $, defined as
\begin{equation}
c\left( \mathbf{r}_{1},\mathbf{\hat{u}}_{1},t\right) =\int_{\mathbf{v}_{1},{\bm\omega }_{1}}f_1\left( \mathbf{r}_{1},\mathbf{v}_{1},\mathbf{\hat{u}}_{1},{%
\bm\omega }_{1},t\right)\;. \label{c}
\end{equation}
Specifically, we seek to derive a kinetic equation
for self propelled particles that is analogous to the mean field
Smoluchowski equation for thermal nematics. In the remainder of this section we present a systematic method for deriving such a closed kinetic equation for $c$ in the limit of low density.

  We adopt
the simplest phenomenological closure of the Fokker-Planck equation \eqref{FP} of the form $f_2
\left( \Gamma _{1},\Gamma _{2},t\right) =f_1\left( \Gamma _{1},t\right)
f_1\left( \Gamma _{2},t\right)$. With this closure, Eq.~\eqref{FP}
 becomes a Boltzmann-Fokker-Planck equation for the one-particle distribution function,
\begin{eqnarray}
\frac{\partial f_1}{\partial t}+\mathcal{D}f _1=\int d\Gamma_2~
\overline{T}\left( 1,2\right) f_1(\Gamma_1,t) f_1(\Gamma_2,t)   \label{BFP}
\end{eqnarray}
To derive the Smoluchowski equation, in addition to the concentration field $c\left( \mathbf{r},\mathbf{\hat{u}}%
,t\right) $ given in Eq.~\eqref{c}, we introduce translational and rotational currents $\mathbf{J}^T$ and $%
{\bf J}^{R}$, defined as velocity moments of the one
particle distribution function, with
\begin{subequations}
\begin{gather}
\mathbf{J}^{T}\left( \mathbf{r}_{1},\mathbf{\hat{u}}_{1},t\right) =\int_{\mathbf{v}_{1},{\bm\omega }_{1}}\mathbf{v}_{1}~f_1\left( \mathbf{r}_{1},\mathbf{%
\hat{u}}_{1},\mathbf{v}_{1},{\bm\omega }_{1},t\right)\;,\\
\mathbf{J}^{R}\left( \mathbf{r}_{1},\mathbf{\hat{u}}_{1},t\right) =\int_{\mathbf{v}_{1},{\bm\omega }_{1}}
{\bm\omega }_{1}~f_1\left( \mathbf{r}_{1},%
\mathbf{\hat{u}}_{1},\mathbf{v}_{1},{\bm\omega }_{1},t\right)\;.
\end{gather}
\end{subequations}
The dynamical equations for the concentration $c$ and the translational and rotational currents are obtained by taking moments of Eq.~\eqref{BFP}, with the result,
\begin{subequations}
\begin{gather}
\frac{\partial c}{\partial t}+\bm\nabla _{\mathbf{r}_{1}}\cdot \mathbf{J}%
^{T}+{\mathbf{\cal R}}_1 \cdot \mathbf{J}^{R}=0  \label{Smolu}\\
\frac{\partial J_{\alpha }^{T}}{\partial t}+\frac{\zeta _{\alpha \beta }^{1}}{m}J_{\beta
}^{T}-\frac{F}{m}\widehat{{u}}_{1\alpha }c+\partial _{1\beta
}\langle v_\alpha v_\beta\rangle c+{\cal { R}}_{1\beta }\langle \omega_\alpha v_\beta\rangle c=-I_{\alpha }^{T}  \label{Flux1}\\
\frac{\partial J_{\alpha }^{R}}{\partial t}+\zeta ^{R}J_{\alpha }^{R}+\partial
_{1\beta }\langle \omega_\alpha v_\beta\rangle c+{\cal { R}}_{1\beta }
\langle \omega_\alpha \omega_\beta\rangle c=-I_{\alpha }^{R}  \label{Flux2}
\end{gather}
\end{subequations}
where
\begin{subequations}
\begin{gather}
\langle v_\alpha v_\beta\rangle=\frac{1}{c}\int_{\mathbf{v}_{1},\bm\omega_{1}}v_{1\alpha
}v_{1\beta }~f_1(\Gamma_1,t)  \label{c1}\\
\langle \omega_\alpha v_\beta\rangle=\frac{1}{c}\int_{\mathbf{v}_{1},\bm\omega_{1}}{\omega }%
_{1\alpha }v_{1\beta }~f_1(\Gamma_1,t)  \label{c2}\\
\langle \omega_\alpha \omega_\beta\rangle=\frac{1}{c}\int_{\mathbf{v}_{1},\bm\omega_{1}}\omega
_{1\alpha }\omega _{1\beta }~f_1(\Gamma_1,t)  \label{c3}
\end{gather}
\end{subequations}
are the second moments of $f_1$ with respect to the translational and rotational
velocities, and
\begin{subequations}
\begin{gather}
I_{\alpha }^{T}=\int d\Gamma_{2}~\left[T\left( 1,2\right)
v_{1\alpha }\right]f_1(\Gamma_1,t)f_1(\Gamma_2,t)  \label{c4}\\
I_{\alpha }^{R}=\int d\Gamma_{2}~\left[T\left( 1,2\right)
\omega _{1\alpha }\right]f_1(\Gamma_1,t)f_1(\Gamma_2,t)   \label{c5}
\end{gather}
\end{subequations}
are  the linear and angular momentum transfers, respectively, due to collisions
with the other particles in the system.

The equations for the translational and rotational currents contain frictional damping and relax on time scales of order $m/\zeta_\Vert$ due to the interaction with the substrate. On time scales $t\gg m\zeta_\Vert^{-1}$
 the fluxes can be approximated as
\begin{subequations}
\begin{gather}
\lim_{t>>m/\zeta_\Vert}J_{\alpha }^{T}=-m(\zeta _{\alpha \beta }^{1})^{-1}\left[ -\frac{F}{m}
\widehat{{u}}_{1\beta }c+\partial _{1\beta
}\langle v_\alpha v_\beta\rangle c+{\cal { R}}_{1\beta }\langle \omega_\alpha v_\beta\rangle c+I_{\alpha }^{T}%
\right]\;,\\
\lim_{t>>\zeta ^{-1}}J_{\alpha }^{R}=-\zeta_{R}^{-1}\left[\partial
_{1\beta }\langle \omega_\alpha v_\beta\rangle c+{\cal { R}}_{1\beta }
\langle \omega_\alpha \omega_\beta\rangle c+I_{\alpha }^{R}%
\right]\;.
\end{gather}\label{Flux_OD}
\end{subequations}
In this regime we can fully describe the system's dynamics in terms of  the one-particle configurational probability $c({\bf r},\hat{\bf u},t)$. To obtain a closed equation, we must express the currents
as functionals of the configurational probability.

\subsection{Thermal Hard Rods: Derivation of the Onsager Excluded Volume Interaction}

Before deriving the Smoluchwski equation for self-propelled hard rods, we show in this section how this method can be implemented to provide a derivation of the Smoluchowski equation for thermal hard rods, with the well-known mean field Onsager excluded volume interaction. We set  the
self-propulsion force $F=0$ and assume the noise is thermal, i.e. $T_a=T$.

For large friction the relaxation of the linear and angular velocities of the rods
is controlled primarily  by the
interaction of the rods with the substrate, rather than by
interparticle collisions. We can then assume that the for times $t>>m/\zeta_\Vert$ the velocity distribution
has relaxed to its
noninteracting value. In the
absence of collisions, the Fokker-Planck equation, Eq.~\eqref{FP}, for the one-particle distribution function takes the form
\begin{eqnarray}
\frac{m}{\zeta _{\Vert }}\left( \partial _{t}+\mathbf{v}_{1}\cdot \nabla _{
\mathbf{r}_{1}}+{\bm\omega }_{1}\cdot  \mathbf{{\cal R}}_{1} \right) f_1&&+\left[
\frac{\zeta _{\bot }}{\zeta _{\Vert }}\delta _{\alpha \beta }-\left( 1-\frac{
\zeta _{\bot }}{\zeta _{\Vert }}\right) \hat{u}_{1\alpha }\hat{u}_{1\beta
}\right] \left( \partial _{v_{1\alpha }}v_{1\beta }+\frac{k_{B}T}{m}\partial
_{v_{1\alpha }}\partial _{v_{1\beta }}\right) f_1  \nonumber \\
&&+\frac{m\zeta ^{R}}{\zeta _{\Vert }}\left( \partial _{{\omega }_{1}} {
\omega }_{1}+\frac{k_{B}T}{I}\partial _{\omega _{1}}^{2}\right) f_1=0\;,
\label{FP-noint}
\end{eqnarray}
where we have used the form of the noise amplitude in Eq.~(\ref{2.2a}-\ref{2.2}) to eliminate it in favor of the temperature $T$. The first term on the left hand side of Eq.~\eqref{FP-noint} can be
neglected for $t>>m/\zeta_\Vert$. The solution of the non-interacting Fokker-Planck equation can then be written in a factorized form as
\begin{equation}
f_1\left( \mathbf{r}_{1},\mathbf{v}_{1},\mathbf{\hat{u}}_{1},{\bm\omega }%
_{1},t\right) =c\left( \mathbf{r}_{1},\mathbf{\hat{u}}_{1},t\right)f_M(v_1,\omega_1)\;,
\end{equation}
with
\begin{equation}
f_M(v_1,\omega_1)=A \exp
\left( -mv_{1}^{2}/2k_{B}T\right) \exp (-I\omega _{1}^{2}/2k_{B}T)\;,
\label{fM}
\end{equation}
and $A$ a normalization constant. In other words, we assume that on the time scales of interest the linear and angular velocity distributions have relaxed to their equilibrium forms.
Using this expression, the velocity moments defined in  Eqs.~(\ref{c1}-\ref{c5}) can be immediately calculated with the result,
\begin{subequations}
\begin{gather}
\langle v_\alpha v_\beta\rangle= \frac{k_{B}T}{m}\delta _{\alpha \beta }\;,\\
\langle v_\alpha \omega_\beta\rangle = 0\;,\\
\langle \omega_\alpha \omega_\beta\rangle =\frac{k_{B}T}{I}\delta _{\alpha \beta }\;.
\end{gather}
\end{subequations}
Next, we need to evaluate the collisional transfer contributions, $I_{\alpha
}^{T}$ and $I_{\alpha }^{R}$ defined in Eqs.~\eqref{c4} and \eqref{c5}.  This requires calculating the mean linear and angular velocity transferred in a collision,  $\left\langle T\left( 1,2\right) v_{1\alpha
}\right\rangle _{M}$ and $\left\langle T\left( 1,2\right) \omega _{1\alpha
}\right\rangle _{M}$ where
\begin{equation}
\left\langle X\right\rangle _{M}=\int_{{\bf v}_{1},\bm\omega _{1}}\int_{{\bf v}_{2},\bm\omega _{2}}X~f_M\left( v_1,\omega_1\right)
f_M\left( v_2,\omega_2\right)\;.
\end{equation}
Using the explicit form of the momentum transfer in a binary hard rod collision given in Eq.~\eqref{A}, we find
\begin{eqnarray}
\left\langle T\left( 1,2\right) \mathbf{v}_{1\alpha }\right\rangle _{M}
&=&-\int_{s_{1}s_{2}\hat{k}}\left\langle \left( \mathbf{V}_{12}\cdot
\widehat{\mathbf{k}}\right) ^{2}\Theta \left( -\mathbf{V}_{12}\cdot \widehat{
\mathbf{k}}\right) \right\rangle _{M}  \nonumber \\
&&\times \frac{1}{1+\frac{m}{2I}\left( \bm{\xi }_{1}\times \mathbf{\hat{k
}}\right) ^{2}+\frac{m}{2I}\left( \bm{\xi }_{2}\times \mathbf{
\hat{k}}\right) ^{2}}\delta \left( \Gamma _{cont}\left( s_{1},s_{2},\widehat{
\mathbf{k}}\right) \right) \hat{k}_{\alpha }\;.  \label{T12M}
\end{eqnarray}
The $\Theta $ function in Eq.~\eqref{T12M} selects only those values of the
pre-collision velocities that will actually result in a collision.  In the mean field limit of interest here, we
assume that on average half the particles in
the flux will have initial velocities  that will yield a collision and approximate the velocity average in Eq.~\eqref{T12M} as
\begin{eqnarray}
\left\langle \left( \mathbf{V}_{12}\cdot \widehat{\mathbf{k}}\right)
^{2}\Theta \left( -\mathbf{V}_{12}\cdot \widehat{\mathbf{k}}\right)
\right\rangle _{M} &\sim &\frac{1}{2}\left\langle \left( \mathbf{V}%
_{12}\cdot \widehat{\mathbf{k}}\right) ^{2}\right\rangle _{M}  \nonumber \\
&=&\frac{1}{2}\left[ \frac{2k_{B}T}{m}+\frac{k_{B}T}{I}\left( \bm{\xi }%
_{1}\times \mathbf{\hat{k}}\right) ^{2}+\frac{k_{B}T}{I}\left( \bm{\xi }%
_{2}\times \mathbf{\hat{k}}\right) ^{2}\right]\;.\label{v2M}
\end{eqnarray}
Substituting Eq.~\eqref{v2M} in Eq.~\eqref{T12M} we find
\begin{equation}
\left\langle T\left( 1,2\right) \mathbf{v}_{1\alpha }\right\rangle _{M}\simeq-%
\frac{k_{B}T}{m}\int_{s_{1}s_{2}}\int_{\hat{\bf k}}\delta \left( \Gamma _{cont}\left(
s_{1},s_{2},\widehat{\mathbf{k}}\right) \right) \hat{k}_{\alpha }\;.\label{T12v}
\end{equation}
Similarly, it is easy to show that
\begin{equation}
\left\langle T\left( 1,2\right) \omega _{1\alpha }\right\rangle _{M}=-\frac{
k_{B}T}{I}\int_{s_{1}s_{2}}\int_{\hat{\bf k}}\delta \left( \Gamma _{cont}\left(
s_{1},s_{2},\widehat{\mathbf{k}}\right) \right) \left( {\bf s }\times
\mathbf{\hat{k}}\right) _{\alpha }\;,\label{T12o}
\end{equation}
where ${\bf s}$ is defined in Fig.~\ref{fig:ExVol}.
In the thin rod limit, i.e., $R\rightarrow
0$, the contact delta function in Eqs.~\eqref{T12v} and \eqref{T12o} is non zero on the
perimeter of a parallelogram centered at the position ${\bf r}_1$ rod $1$ with  sides of
length $\ell $ directed along $\mathbf{\hat{u}}_{1}$ and
$\mathbf{\hat{u}}_{2}$ (Fig.~\ref{fig:ExVol}). The area  of such a parallelogram
can be written as
\begin{equation}
A_{\Vert gram}({\bf r}_1,\hat{\bf u}_1)=\ell^2\int_{s_{1},s_{2}}\Theta \left( 0^{+}-\left| {\bf r}_{12}+s_{1}%
\hat{\bf u}_{1}-s_{2}\hat{\bf u}_{2}\right| \right)\;.\label{area}
\end{equation}
It is easy to verify that
\begin{subequations}
\begin{gather}
\bm\nabla _{\mathbf{r}_{1}}A_{\Vert gram}=-\ell^2\int_{s_{1},s_{2}}\int_{\hat{\bf k}_{21}}\delta
\left( \Gamma _{cont}\left( s_{1},s_{2},\hat{\mathbf{k}}_{21}\right) \right)
\mathbf{\hat{k}_{21}}\;,\\
{\cal R}_1 A_{\Vert gram}=-\ell^2\int_{s_{1},s_{2}}\int_{\hat{\bf k}_{21}}\delta \left(
\Gamma _{cont}\left( s_{1},s_{2},\widehat{\mathbf{k}}\right) \right) \left( {%
\bf s }\times \mathbf{\hat{k}_{21}}\right)\;.
\end{gather}
\end{subequations}
Finally, combining all these results we obtain the familiar
Smoluchowski equation for hard given in Eqs.~\eqref{Smol0},~(\ref{JR}) and (\ref{JT}) with
%
%
$D_{\alpha \beta }=\zeta _{\alpha \beta }^{-1}k_{B}T/m$ and $D_{R}=k_{B}T/(\zeta_RI)$. The excluded volume interaction can be written in the familiar Onsager form by noting that the area of the collisional parallelogram is simply given by $A^{\Vert gram}=\ell^2|\hat{\bf u}_1\times\hat{\bf u}_2|$ as illustrated in Fig.~(\ref{fig:ExVol}). This gives Eq.~\eqref{Vex}, where
 the integral over ${\bf s}$ spans the area of the parallelogram shown in Fig.~(\ref{fig:ExVol}) for fixed orientation of rod $2$, while the integral over $\bf{\hat{u}}_{2}$ averages over all possible orientations of the second rod. The Onsager excluded volume interaction has a
purely entropic interpretation as the free energy cost for
a rod to occupy the area excluded by another rod.
Here the same result has been derived from
dynamical considerations and has the alternate interpretation
of mean-field momentum transfer in a binary collision of two rods, each  carrying an average momentum of magnitude $\sqrt{mk_{B}T}$. The orientational
correlations arise then from the anisotropy of the collision
frequency due to the fact  that head to head collisions are of measure zero with respect to
head to side collisions.

In summary, we have shown in this subsection that for thermal (non self-propelled) hard rods the method described in this paper can be used to derive the familiar Smoluchowski equation  with the Onsager expression for the excluded volume interaction. In the next section we apply the same procedure to self-propelled hard rods and show that  self-propulsion modifies the Smoluchowski equation in several ways.

\subsection{Self-propelled hard rods}

As for the case of thermal hard rods, the derivation of the Smoluchowski equation for self propelled rods consists of two steps: (i) we identify a stationary normal solution of
the noninteracting Boltzmann-Fokker Plank equation, and (ii) we use this functional
form to close the equations for the translational and rotational fluxes in a
quasi-static approximation.

In the absence of interactions, the Fokker-Planck equation for self-propelled rods is given by
\begin{gather}
\frac{m}{\zeta _{\Vert }}\left( \partial _{t}+\mathbf{v}_{1}\cdot \nabla _{
\mathbf{r}_{1}}+\bm{\omega} _{1}\cdot{\bf{\cal R}}_{1}\right) f_1+v_{0}\mathbf{
\hat{u}}_{1}\mathbf{\cdot \nabla }_{\mathbf{v}_{1}}+\left[
\frac{\zeta _{\bot }}{\zeta _{\Vert }}\delta _{\alpha \beta }-\left( 1-\frac{
\zeta _{\bot }}{\zeta _{\Vert }}\right) \hat{u}_{1\alpha }\hat{u}_{1\beta
}\right] \left( \partial _{v_{1\alpha }}v_{1\beta }+\frac{k_{B}T}{m}\partial
_{v_{1\alpha }}\partial _{v_{1\beta }}\right) f _1 \nonumber \\
+\frac{m\zeta ^{R}}{\zeta _{\Vert }}\left( \partial _{{\omega }_{1}} {
\omega }_{1}+\frac{k_{B}T}{I}\partial _{\omega _{1}}^{2}\right) f_1=0\;\label{nonintSP}
\end{gather}
with $v_{0}=\frac{F}{\zeta _{\Vert }}$ the self propulsion velocity. On time scales large compared to $m/\zeta_\Vert$, we neglect the  first term on the left hand side of Eq.~\eqref{nonintSP}. The stationary normal solution then has  the form
\begin{equation}
f_1\left( \mathbf{r}_{1},\mathbf{\hat{u}}_{1},\mathbf{v}_{1},{\bm\omega }%
_{1},t\right) =c\left( \mathbf{r}_{1},\mathbf{\hat{u}}_{1},t\right)
f_S \left( \mathbf{v}_{1},{\bm\omega }_{1}|\mathbf{\hat{u}}%
_{1}\right)\;,
\end{equation}
with
\begin{equation}
f_S=A\exp \left( -\frac{1}{2k_{B}T}m\left( {\bf v}_{1}-v_{0}%
\widehat{\mathbf{u}}_{1}\right) ^{2}-\frac{1}{2k_{B}T}I\omega
_{1}^{2}\right) \;,
\end{equation}
and $A$ a normalization constant.
Inserting this stationary normal solution in Eqs.~\eqref{Flux_OD},  we obtain
\begin{subequations}
\begin{gather}
J_{\alpha }^{T} =v_{0}\widehat{{u}}_{1\alpha }c-D_{S}\hat{u}
_{1\alpha }\mathbf{\hat{u}}_{1}\mathbf{\cdot \nabla }_{\mathbf{r}
_{1}}c-D_{\alpha \beta }\partial _{\mathbf{r}_{1\beta }}c-m\zeta _{\alpha \beta }^{-1}\left( \int d\mathbf{r}_{2}d\mathbf{\hat{u}}
_{2}\left\langle T\left( 1,2\right) \mathbf{v}_{1\beta }\right\rangle
_{S}c\left( \mathbf{r}_{2},\mathbf{\hat{u}}_{2},t\right) \right)
 c\left( \mathbf{r}_{1},\mathbf{\hat{u}}_{1},t\right)\;, \label{JTsp_1}\\
\mathbf{J}^{R} =-D_{R}{\bf {\cal R}}_{1} c  -\frac{1}{\zeta ^{R}}\left( \int d\mathbf{r}_{2}d\mathbf{\hat{u}}
_{2}\left\langle T\left( 1,2\right) {\bm\omega }_{1}\right\rangle
_{S}c\left( \mathbf{r}_{2},\mathbf{\hat{u}}_{2},t\right) \right)   c\left( \mathbf{r}_{1},\mathbf{\hat{u}}_{1},t\right)\;,\label{JRsp_1}
\end{gather}
\end{subequations}
where the angular brackets $\langle ...\rangle_S$  denote the average over the linear and angular velocities
with weigh  $f_S$. The diffusion coefficients $
D_{\alpha \beta }$ and $D_{R}$ are as given in the previous subsection, but with $T=T_a$, and $D_{S}=
\frac{v_{0}^{2}}{\zeta _{\Vert }}$.

The next step  is the evaluation of  the
average linear and angular momentum transfer in a collision,
\begin{subequations}
\begin{gather}
\left\langle T\left(1,2\right) \mathbf{v}_{1}\right\rangle _{S}=\int_{{\bf v}_1,\bm\omega_1}\int_{{\bf v}_2,\bm\omega_2}
T(1,2) {\bf v}_1~f_S({\bf v}_1,\bm\omega_1|{\mathbf{\hat u}}_1)f_S({\bf v}_2,\bm\omega_2|{\mathbf{\hat u}}_2)\;,\\
\left\langle T\left(
1,2\right) {\bm\omega }_{1}\right\rangle _{S} =\int_{{\bf v}_1,\bm\omega_1}\int_{{\bf v}_2,\bm\omega_2}
T(1,2) \bm\omega_1~f_S({\bf v}_1,\bm\omega_1|{\mathbf{\hat u}}_1)f_S({\bf v}_2,\bm\omega_2|{\mathbf{\hat u}}_2)\;.
\end{gather}
\end{subequations}
When deriving the
Onsager excluded volume interaction in the previous subsection, we neglected
dynamical velocity correlations by  approximating $
\left\langle ( \mathbf{V}_{12}\cdot \widehat{\mathbf{k}})
^{2}\Theta ( -\mathbf{V}_{12}\cdot \widehat{\mathbf{k}})
\right\rangle \sim \frac{1}{2}\left\langle ( \mathbf{V}_{12}\cdot
\widehat{\mathbf{k}}) ^{2}\right\rangle $. This approximation is not valid for self-propelled rods
as  the dynamical velocity correlations induced by the self
propulsion velocity, which is directed along $\mathbf{
\hat{u}}$, are also \textit{orientational} correlations.
On the other hand, an approximation is required to make the problem  tractable.

The Smoluchowski equation  is a mean
field model and only describes the average momentum transfer in a collision.
We then assume
\begin{subequations}
\begin{gather}
\left\langle
T\left( 1,2\right) \mathbf{v}_{1 }\right\rangle _{S}\simeq \left\langle
T\left( 1,2\right) \mathbf{v}_{1 }\right\rangle _{M_a}+\left\langle
T\left( 1,2\right) \mathbf{v}_{1 }\right\rangle _{SP}\;,\\
\left\langle
T\left( 1,2\right) \bm\omega_{1 }\right\rangle _{S}\simeq \left\langle
T\left( 1,2\right) \bm\omega_{1 }\right\rangle _{M_a}+\left\langle
T\left( 1,2\right)\bm\omega_{1 }\right\rangle _{SP}\;,
\end{gather}\label{Tapprox}
\end{subequations}
where $\left\langle X\right\rangle _{M_a}$ denotes velocity averages with the Maxwellian
distribution given in Eq.~\eqref{fM} at temperature $T=T_a$  and $\left\langle
X\right\rangle _{SP}$ denotes velocity averages in a regime where all particles are moving at velocity
$v_0\mathbf{\hat u}$ and  the one-rod velocity distribution is given by $f_{SP }\sim \delta \left(
\mathbf{v}-v_{0}\mathbf{\hat{u}}\right) \delta \left( \bm\omega \right) $. The first terms on the right hand side of Eqs.~\eqref{Tapprox} are then evaluated as in the previous section  by
neglecting the dynamical velocity correlations and yield again the
Onsager excluded volume potential given in Eq.~\eqref{Vex}. The second terms on the right hand side of Eqs.~\eqref{Tapprox} are evaluated by neglecting static orientational correlations
arising from the term  $\left[1+(m/2I)( \bm{\xi }_{1}\times \mathbf{\hat{%
k}}) ^{2}\mathbf{+}(m/2I)( \bm{\xi }_{2}\times \mathbf{%
\hat{k}}) ^{2}\right]^{-1}$  in Eq.~(\ref{A}) as these correlations are already incorporated  in the
noise. It can be shown that this approximation becomes exact if we replace the rod by
a string of beads in contact with each other and calculate the momentum
transfer between  the two specific beads  that participate in the collision. With these
approximations, the average momentum transfer is given by
\begin{subequations}
\begin{gather}
\left( \int_{2} \left\langle T\left( 1,2\right) \mathbf{v}_{1\beta }\right\rangle
_{S}c\left( 2 \right)\right)
 c\left( 1 \right)\simeq
\frac{1}{m}\left( \nabla _{\mathbf{r}_{1\alpha }}V_{ex}\right)
c+v_{0}^{2}I_{\alpha }^{SP}\;\\
\left( \int_{2} \left\langle T\left( 1,2\right) {\bm\omega }_{1}\right\rangle
_{S}c\left( 2\right) \right)   c\left( 1\right) \simeq
\frac{1}{I}{\bf {\cal R}}_{1} V_{ex}c+\frac{v_{0}^{2}}{2I}I_{R}^{SP}\;,
\end{gather}
\end{subequations}
with
\begin{eqnarray}
I^{SP} &=&\int_s\int_{\mathbf{\hat{u}}_{2}}\sin ^{2}\left(
\theta _{1}-\theta _{2}\right) \left[ \Theta \left( \sin \left( \theta
_{1}-\theta _{2}\right) \right) -\Theta \left( -\sin \left( \theta
_{1}-\theta _{2}\right) \right) \right]   \nonumber \\
&&\times \left[ \mathbf{\hat{u}}_{1}^{\bot }~c\left( \mathbf{r}_{1}+s\mathbf{%
\hat{u}}_{1}-\frac{\ell}{2}\mathbf{\hat{u}}_{2},\mathbf{\hat{u}}_{2},t\right) +%
\mathbf{\hat{u}}_{2}^{\bot }~c\left( \mathbf{r}_{1}+s\mathbf{\hat{u}}_{2}-%
\frac{\ell}{2}\mathbf{\hat{u}}_{1},\mathbf{\hat{u}}_{2},t\right) \right]\;,
\label{SP_trans}
\end{eqnarray}
\begin{eqnarray}
\mathbf{I}_{R}^{SP} &=&\mathbf{\hat{z}}\int_{s}\int_{\mathbf{%
\hat{u}}_{2}}\sin ^{2}\left( \theta _{1}-\theta _{2}\right) \left[ \Theta
\left( \sin \left( \theta _{1}-\theta _{2}\right) \right) -\Theta \left(
-\sin \left( \theta _{1}-\theta _{2}\right) \right) \right]   \nonumber \\
&&\times \left[ s~c\left( \mathbf{r}_{1}+s\mathbf{\hat{u}}_{1}-\frac{\ell}{2}%
\mathbf{\hat{u}}_{2},\mathbf{\hat{u}}_{2},t\right) +\frac{\ell}{2}\cos \left(
\theta _{1}-\theta _{2}\right) c\left( r_{1}+s\mathbf{\hat{u}}_{2}-\frac{\ell}{2%
}\mathbf{\hat{u}}_{1},\mathbf{\hat{u}}_{2},t\right) \right] \;.  \label{SP_rot}
\end{eqnarray}
Finally, when these results are substituted into Eq.~(\ref{JTsp_1}) and (\ref{JRsp_1}), we obtain the modified
Smoluchowski equation given in Eq.~(\ref{JSP}).

There are three important modifications of the Smoluchowski
equation for self-propelled particles when compared to its equilibrium
counterpart for thermal particles.
\begin{enumerate}
\item The  translational current $\mathbf{J}^{T}$ in Eq.~(\ref
{JTsp}) contains a convective term $v_{0}\widehat{\mathbf{u}}_{1 }c
$ that arises because self-propelled particles move in the
direction of their long axis.  This is a signature of the polar nature
of the microdynamics of the system.

\item  Self propulsion yields  an additional longitudinal diffusion current $D_{S}%
\hat{u}_{1\alpha }\hat{u}_{1\beta}\partial_{1\beta}c$,
with $D_{S}\sim v_{0}^{2}$. This can be understood as follows.
A Brownian particle subject to a frictional damping $\zeta$ takes Brownian steps of mean length
$\Delta\sim\sqrt{mk_BT}/\zeta$,  with $\sqrt{k_{B}T/m}$ the thermal
speed  of the particle and $m/\zeta$ the frictional time scale over which the velocity relaxes to zero. This yields the simple estimate
$D\sim \Delta^2/(m/\zeta)\sim k_BT/\zeta$ for the diffusion coefficient. A Brownian rod experiences anisotropic friction $\zeta_\Vert$ and $\zeta_\bot$, yielding mean steps $\Delta_\Vert\sim\sqrt{mk_BT}/\zeta_\Vert$  and
$\Delta_\bot\sim\sqrt{mk_BT}/\zeta_\bot$ in the directions longitudinal and transverse to its long axis, respectively. This gives anisotropic diffusion constants $D_\Vert\sim k_BT\zeta_\Vert$ and $D_\bot\sim k_BT/\zeta_\bot$. When the rod is self
propelled it performs a persistent random walk along its long axis controlled by the competition between ballistic motion at speed $v_0$ and rotational diffusion at rate $D_R\sim D_\Vert/\ell^2$. As a result, the mean square velocity is approximately $\langle v^2\rangle\sim \frac{k_BT}{m}+v_0^2$. For small values of $v_0$ this gives
$\langle v^2\rangle^{1/2}\sim\frac{k_BT}{m}\left[1+\frac{mv_0^2}{2k_BT}\right]$. This yields an additional contribution of order $v_0^2$ to the longitudinal diffusion, corresponding to
the second term of Eq.~(\ref{DSP}).

\item Both the translational and rotational fluxes in Eq.~(\ref{JSP}) contain additional active contributions arising from the momentum transfer associated with self propulsion.  Collisions among particle induce
dynamical velocity correlations. These are negligible for a mean-field description of
overdamped thermal hard rods that only seeks to capture the dynamics of the translational and orientational degrees of freedom. When the rods are self-propelled  with a physical velocity directed along
their long axis, collisions induce {\em both} velocity and orientational correlations since the two are intimately coupled. These additional collision-induced orientational
correlations have been shown to affect the physics of the system even on hydrodynamic scales~\cite{ABMCM_PRL}.
\end{enumerate}

Our work demonstrates that  orientational fluctuations have a profound effect on mass
transport in  self-propelled particle systems. Furthermore this observation is not limited to the
specific hard rod model considered here, but holds generically for all collections of self-propelled units.  Our result relies solely on the presence of short-range excluded volume interactions that are present in all physical systems, but not on the specific implementation of such interaction in the hard particle limit.

\section{Discussion}

In this paper we have analyzed the microscopic dynamics and statistical mechanics of a
collection of  self-propelled particles modeled as long thin
polar rods that move along one direction of their long axis.  The formalism
developed here is general and of wide applicability. It can be used to study
fluctuations and response functions in fluids of self-propelled particles  by building on techniques developed for traditional fluids.
In addition, the formalism
 can readily be generalized to particles of arbitrary
shape and to higher dimensions, making it also relevant for applications to
 granular fluids.

As a particular application of the general formalism, we have
derived the Smoluchowski equation that governs the dynamics of the
one particle configuration probability density in the  overdamped regime, for both thermal and
self propelled particles. For thermal rods,
this provides a purely dynamical derivation of the familiar
Onsager excluded volume interaction and  is useful for
identifying the limitations of this widely used effective interaction in capturing the physics of out-of-equilibrium systems. For self
propelled hard rods, the modified Smoluchowski equation presented
here is the first tractable theory of the dynamics of
self-propelled particles that captures the physics of contact
interactions and their modifications due to the presence of self
propulsion. In a separate work we have used this  Smoluchowski equation
as the starting point for deriving a coarse grained (hydrodynamic) description of the system in terms of
conserved quantities and broken symmetry variables~\cite{ABMCM_PRL,Aparna1}.
Self propulsion has profound effects on the system on hydrodynamic scales,
including  enhanced orientational
order arising from the  orientational correlations induced
dynamically via collisions, instabilities of the ordered phases,
 and the existence of propagating sound-like density waves in this otherwise
overdamped system.

Having derived the Smoluchowski equation from first principles, we
can also identify its scope and limitations. Our theory captures the
fluctuations in velocity induced by orientational fluctuations and
the associated modifications to the mass flux that characterize
this inherently nonequilibrium system. These effects yield
the convective and diffusive terms in Eq.~(\ref{JSP}).
The theory also captures some of the orientational correlations induced
by collisional interactions through the additional momentum
transfer contributions to the fluxes. These correlations are
important as evidenced by the enhanced ordering identified in
\cite{ABMCM_PRL} and observed in numerical simulations of
motility assays \cite{Lipowsky}. The derivation is, however,
based on a low-density kinetic theory that neglects two-particle velocity correlations.
 While this is a reasonable approximation for overdamped
thermal particles that have an underlying equilibrium state,
in the case of self propelled particles, these correlations will
generate additional orientational correlations that are neglected in the present approximation.
The Fokker-Planck hierarchy  derived here
can serve as  the starting point for analyzing the effect of these two-particle correlations.
This is left for future work.

In summary, we have constructed the non-equilibrium statistical mechanics of
a system of self propelled particles and derived a modified Smoluchowski
equation for the system. We have discussed the content
of the resulting theory and identified its scope, limitations and potential
for future applications.

\begin{acknowledgements}
This work  was supported by NSF grants DMR-075105 and DMR-0806511.
\end{acknowledgements}


\begin{thebibliography}{99}

\bibitem{TonerPRL} J. Toner and Y. Tu, Phys. Rev. Lett. {\bf 75}, 4326 (1995).

\bibitem{TonerRev}  J. Toner, Y. Tu and S. Ramaswamy, Ann. Phys. {\bf 318}, 170 (2005).

\bibitem{KruseEPJE2005}
K. Kruse, J. F. Joanny, F. J\"ulicher, J. Prost, and K. Sekimoto,
Eur. Phys. J. E {\bf 16}, 5Ð16 (2005).

\bibitem{Orsogna} M. R. D'Orsogna, Y. L. Chuang, A. L. Bertozzi and L. S. Chayes, Phys. Rev. Lett.{\bf 96}, 104302 (2006).

\bibitem{Chate2} H. Chat\'e, F. Ginelli and R. Montagne, Phys. Rev.
Lett., {\bf 96}, 180602 (2006).

\bibitem{Lipowsky}   P. Kraikivski, R. Lipowsky and J. Kierfeld, Phys. Rev. Lett. {\bf 96}, 258103
(2006).

\bibitem{bactexpts} O. A. Igoshin, R. Welch and D. Kaiser, PNAS, {\bf 101}, 4256
(2004).

\bibitem{actinexpts}F. J. N\'ed\'elec, T. Surrey, A. C. Maggs and S. Leibler, Nature {\bf 389}, 305
(1997).

\bibitem{rodsexpt1}V. Narayanan, S. Ramaswamy and N. Menon, Science
{\bf 317}, 105 (2007).

\bibitem{rodsexpt2}A. Kudrolli, G. Lumay, D. Volfson and L. S.
Tsimring, Phys. Rev. Lett. {\bf 100}, 058001 (2008).

\bibitem{experiments}  A. Sokolov, I. S. Aranson, J. O. Kessler and R. E. Goldstein, Phys. Rev. Lett. {\bf 98},158102 (2007).

\bibitem{ABMCM_PRL} A. Baskaran and M. C. Marchetti, Phys. Rev.
Lett., {\bf 101}, 268101 (2008).

\bibitem{Aparna1} A. Baskaran and M. C. Marchetti, Phys. Rev. E
{\bf 77}, 011920 (2008).

\bibitem{EdDoiBook} M Doi, S. F Edwards, {\it The Theory of Polymer Dynamics}, Oxford University Press (1986).

\bibitem{Huthmann1999}
M. Huthmann, T. Aspelmeier, and A. Zippelius, Phys. Rev. E {\bf 60}, 654 (1999).

\bibitem{Otto2006}
M. Otto, T. Aspelmeier and  A. Zippelius, J. Chem. Phys. {\bf 124}, 154907 (2006).

\bibitem{OsendalBook} C. W. Gardiner, {\it Handbook of Stochastic
Methods}, Springer, New York (1983).

\bibitem{Onsager} L. Onsager,
Annals of New York Academy of Sciences, {\bf 51}, 627 (1949).

\bibitem{PersistentRW1}  F. Peruani, L. G.
Morelli, Phys. Rev. Lett, {\bf 99}, 010602 (2007).

\bibitem{PersistentRW2} D. Selmeczi, S. Mosler, P. H. Hagedorn, N. B. Larsen, and H.
Flyvbjerg, Biophys. J. {\bf 89}, 912 (2005).

\end{thebibliography}
\end{document}